\documentclass[12pt,a4paper]{article}
\usepackage{cmap}
\usepackage[T2A]{fontenc}
\usepackage[cp1251]{inputenc}
\usepackage[english, russian]{babel}
\usepackage{amsfonts,amsmath,amssymb,amsthm,longtable,hhline}
\usepackage{bm}
\usepackage{cite}
\usepackage{color,soul}
\usepackage{color,xcolor}
\usepackage{multicol}
\usepackage{tabularx,caption,multirow}
\usepackage{ulem}
\usepackage{titlesec}

\makeatletter
\renewcommand{\@biblabel}[1]{#1.} 
\makeatother
\makeatletter

\newlength\aptextwidth
\definecolor{BrickRed}{rgb}{0.588,0.098,0.055}

\def\noblue#1{\ifmmode \text{#1}\else #1\fi}
\def\noclh#1{\ifmmode \text{#1}\else #1\fi}
\def\rem#1{}

\def\ttable#1. #2{\begin{table}[t]\tablehat{#1}{#2}}
\def\mtable#1. #2{\begin{table}[hbtp]\tablehat{#1}{#2}}
\def\ptable#1. #2{\begin{table}[p]\tablehat{#1}{#2}}
\def\tablehat#1#2{\centering\small \vbox{\parindent=0pt
  \leftskip=0pt plus.5\hsize \rightskip=\leftskip \parfillskip=0pt
  ТАБЛИЦА #1\\ #2}\nobreak\medskip\medskip }
\def\texendtable{\end{table}}
\def\notation{\par\ifnum\lastpenalty<25000 \bigbreak \fi
  \noindent\triangle\enspace\ignorespaces}

\def\hline{\noalign{\hrule}}
\def\rlineskip{\vruleskip\hline\vruleskip}

\let\ds=\displaystyle

\let\bl=\bigl \let\br=\bigr
\let\Bl=\Bigl \let\Br=\Bigr

\def\eqnitemskip{\ifhmode \else \par
  \ifnum\lastpenalty>24999
    \ifnum\lastpenalty=25004 \fi
  \else \medbreak \fi \fi }
\def\eqnitem #1. {\eqnitemskip
  {\setbox0=\hbox{$#1^\circ$.\enspace}%
  \ifdim\wd0>\parindent \box0\ignorespaces \else
  \hbox to\parindent{\unhbox0\hss}\ignorespaces\fi}}
\def\eqnitemnobreak #1. {\noindent
  {\setbox0=\hbox{$#1^\circ$.\enspace}%
  \ifdim\wd0>\parindent \box0\ignorespaces \else
  \hbox to\parindent{\unhbox0\hss}\ignorespaces\fi}}
\newdimen\eqnparindent
\eqnparindent=\parindent
\def\eqnitem #1. {\eqnitemskip\noindent\hskip\eqnparindent $#1^\circ$.\enspace\ignorespaces }
\def\eqnitemnobreak #1. {\noindent\hskip\eqnparindent $#1^\circ$.\enspace\ignorespaces }
\def\simpleitem #1. {\eqnitemskip\noindent\hskip\eqnparindent #1.\enspace\ignorespaces }

\def\eqalignno#1{\displ@y \tabskip\centering
  \halign to\displaywidth{\hfil$\@lign\displaystyle{##}$\tabskip\z@skip
    &$\@lign\displaystyle{{}##}$\hfil\tabskip\centering
    &\llap{$\@lign\eqnofont##$}\tabskip\z@skip\crcr
    #1\crcr}}
\let\eqalignno=\eqalignm
\def\eqcenter#1{\displ@y \tabskip\centering
  \halign{\hfil$\displaystyle{##}$\hfil\crcr
    #1\crcr}}
\def\eqcenterno#1{\displ@y \tabskip\centering
  \halign to\displaywidth{\hfil$\@lign\displaystyle{##}$\hfil
    \tabskip\centering&\llap{$\@lign\eqnofont##$}\tabskip\z@skip\crcr
    #1\crcr}}
\def\texcases#1{\left\{\,\vcenter{\normalbaselines\m@th
    \ialign{$##\hfil$&\quad##\hfil\crcr#1\crcr}}\right.}
\def\Displaylines#1{\vcenter{\displ@y \tabskip\z@skip
  \halign{\hbox to\displaywidth{$\@lign\hfil\displaystyle##\hfil$}\crcr
    #1\crcr}}}

\def\tan{\mathop{\operator@font tg}\nolimits}

\makeatother

\titlespacing\section{0pt}{12pt plus 4pt minus 2pt}{0pt plus 2pt minus 2pt}
\titlespacing\subsection{0pt}{12pt plus 4pt minus 2pt}{0pt plus 2pt minus 2pt}

\begin{document}
\large 

\bigskip

\centerline{\bf\Large Principle of structural analogy of solutions and}
\centerline{\bf\Large its application to nonlinear PDEs and delay PDEs}

\bigskip

\centerline{Andrei D. Polyanin}
 \medskip
\centerline{Ishlinsky Institute for Problems in Mechanics, Russian Academy of Sciences,}
\centerline{pr. Vernadskogo 101, bldg. 1, Moscow, 119526 Russia}
\smallskip
\centerline{e-mail: polyanin@ipmnet.ru}

\bigskip
Using the principle of structural analogy of solutions, approaches have been developed for constructing exact solutions of complex nonlinear PDEs,
including PDEs with delay, based on the use of special solutions to auxiliary simpler related equations.
It is shown that to obtain exact solutions of nonlinear non-autonomous PDEs, the coefficients of which depend on time, it is possible to use
generalized and functional separable solutions of simpler autonomous PDEs, the coefficients of which do not depend on time.
Specific examples of constructing exact solutions to nonlinear PDEs, the coefficients of which depend arbitrarily on time, are considered.
It has been discovered that generalized and functional separable solutions of nonlinear PDEs
with constant delay can be used to construct exact solutions of more complex nonlinear PDEs with variable delay of general form.
A number of nonlinear reaction-diffusion type PDEs with variable delay are described, which allow exact solutions with generalized separation of variables.
\medskip

\textsl{Keywords\/}:
nonlinear PDEs, delay PDEs,
reaction-diffusion equations,
solution methods,
exact solutions,
generalized separable solutions,
functional separable solutions
\medskip

\textsl{Mathematics Subject Classification\/}: 35C05, 35K55, 35K57, 35R10

\section{Introduction. Exact solutions}

Exact solutions of mathematical physics equations and other partial differential equations contribute to a better understanding of qualitative
features of many nonlinear phenomena and processes in various fields of natural science.
Since exact solutions are rigorous mathematical standards,
they can be used as test problems to assess the accuracy and verify the adequacy
of various numerical and approximate analytical methods for solving nonlinear PDEs as well as more complex PDEs with delay.

Exact solutions of nonlinear PDEs are usually understood as solutions that are expressed \cite{polzhu2022}:

(i) through elementary functions or in the form of quadratures (i.e., using elementary functions and indefinite integrals);

(ii) through solutions of ordinary differential equations or systems of such equations.

For more complex PDEs with delay to exact solutions from Items (i) and (ii) it is necessary to add solutions that are expressed through solutions of
ODEs with delay or solutions of ODE systems with delay \cite{polsorzhu2023}.

Exact solutions of nonlinear PDEs are most often constructed using
the classical method of symmetry reductions \cite{ovs1982,blu1989,ibr1994,olv2000},
the direct method of symmetry reductions \cite{polzhu2022,cla1989,hood2000,polzai2012},
the nonclassical symmetries methods \cite{blu1969,lev1989,arr1993,cher2017},
methods of generalized separation of variables \cite{polzhu2022,polzai2012,gal2007,kos2020,kos2022},
methods of functional separation of variables \cite{polzhu2022,polzai2012,doy1998,est2002},
the method of differential constraints \cite{polzhu2022,polzai2012,kap2003,mel2005},
and some other exact analytical methods (see, for example, \cite{kud2005,kud2010,con2020,cal1982,abl1991,aks2021,bed2022}).
On methods for constructing exact solutions of nonlinear delay PDEs and functional PDEs, see, for example,
\cite{polsorzhu2023,mel2008,long2016,polsor2021a,polsor2021b,polsor2021c,aib2021,sor2022}.

This article will describe new approaches for constructing exact solutions of nonlinear non-autonomous equations of mathematical physics and other PDEs
(including PDEs with delay), based on the use of auxiliary solutions to simpler equations.
These approaches are based on the principle of structural analogy of solutions,
which is formulated as follows:
\textit{exact solutions of simpler equations can serve as the basis for constructing solutions to more complex related equations}.
Note that this principle at the heuristic level was successfully used in works \cite{polsorzhu2023,aks2021,polsor2021a,polsor2021b,polsor2021c,polsor2023} to construct exact solutions of nonlinear PDEs with delay,
as well as functional PDEs, using solutions of simpler related PDEs without delay.
In this article, the results obtained using the principle of analogies and illustrated with specific examples,
it was possible to generalize and strictly prove some constructive statements.

\section{General formulations of the problems under consideration}

Let us give general formulations of two important problems discussed in this article.

\textbf{Problem 1.}
\textit{Let the exact solution $u=u(x,t)$ of some nonlinear PDE be known, which depends on the free parameters $a_1,\dots,a_p$.
The problematic question arises: in what cases can we say anything about the exact solutions of a more complex PDE, which is obtained
from the original one by replacing the free parameters with arbitrary functions $a_1(t),\dots,a_p(t)$?}

The next section will show how exact solutions with generalized or functional separation of variables of nonlinear autonomous PDEs,
the coefficients of which do not depend on time, can be used to construct exact solutions of more general nonlinear non-autonomous PDEs,
the coefficients of which depend arbitrarily on time $t$.

\textbf{Problem 2.}
\textit{Let the exact solution $u=u(x,t)$ of some nonlinear PDE with a constant delay $\tau$ be known.
The problematic question arises: in what cases can we say anything about the exact solutions of a more complex PDE, which is obtained
from the original one by replacing the constant delay $\tau$ with an arbitrary variable delay $\tau(t)$?}

Next, in Section~\ref{sec-4} we will show how exact solutions with generalized or functional separation of variables
of nonlinear PDEs with constant delay can be used to construct exact solutions of more general nonlinear PDEs
with variable delay of the general form.

\section{Using solutions of autonomous PDEs to construct exact solutions of non-autonomous PDEs}

Let us describe a method for constructing exact solutions of non-autonomous PDEs, whose coefficients depend on time $t$,
which is based on the use of solutions of simpler autonomous PDEs,
the coefficients of which do not depend on time.
Following the principle ``from simple to complex'', we will first demonstrate the sequence and logic of reasoning in such cases using several specific illustrative examples,
after analyzing which, we will then formulate general conclusions in the form of rigorous statements.

\textit{Example 1}.
Let us consider the nonlinear reaction-diffusion type equation
\begin{equation}
u_t=a_1uu_{xx}+a_2,
\label{xeq12:01}
\end{equation}
where $a_1$ and $a_2$ are free parameters.
Equation \eqref{xeq12:01} admits an exact solution with generalized separation of variables, quadratic in the spatial variable
\begin{equation}
u=\psi_1(t)x^2+\psi_2(t),
\label{xeq12:02}
\end{equation}
where the functions $\psi_1=\psi_1(t)$ and  $\psi_2=\psi_2(t)$ satisfy a nonlinear ODE system,
\begin{equation}
\begin{aligned}
&\psi_1'=2a_1\psi_1^2,\\
&\psi_2'=2a_1\psi_1\psi_2+a_2.
\label{xeq12:03}
\end{aligned}
\end{equation}
Here and below, the primes denote derivatives with respect to $t$.
System \eqref{xeq12:03} is easily integrated, but we will not need this further.

Now formally replacing the parameters $a_1$ and $a_2$ in equation \eqref{xeq12:01} with arbitrary functions $a_1(t)$ and $a_2(t)$, we obtain
\begin{equation}
u_t=a_1(t)uu_{xx}+a_2(t).
\label{xeq12:04}
\end{equation}
It is easy to check that equation \eqref{xeq12:04} also admits an exact solution of the form \eqref{xeq12:02},
where the functions $\psi_1=\psi_1(t)$ and $\psi_2=\psi_2(t)$ satisfy the nonlinear ODE system \eqref{xeq12:03},
in which the parameters $a_1$ and $a_2$ must be respectively replaced by arbitrary functions $a_1(t)$ and $a_2(t)$.

The considered illustrative example allows us to make a simple generalization, which can be formulated in the form of the following statement.
\medskip

\textbf{Statement 1}.
\textit{Let the autonomous nonlinear partial differential equation
	\begin{equation}
		u_t=F\bigl(x,u,u_x,\dots,u^{(n)}_x;a_1,\dots,a_p\bigr),
		\label{bla1*}
	\end{equation}
depending on the free parameters $a_1,\dots,a_p$, has a generalized separable solution in the form of a polynomial in the spatial variable~$x$:
	\begin{align}
		u=\sum^m_{k=0}\psi_k(t)x^k.
		\label{bla2*}
	\end{align}
Then the more complex non-autonomous partial differential equation
	\begin{equation}
		u_t=F\bigl(x,u,u_x,\dots,u^{(n)}_x;a_1(t),\dots,a_p(t)\bigr),
		\label{bla3*}
	\end{equation}
which is obtained from equation \eqref{bla1*} by formally replacing the parameters $a_1,\dots,a_p$ with arbitrary functions $a_1(t),\dots,a_p(t)$,
also admits an exact solution of the form \eqref{bla2*} with other functions $\psi_k(t)$.}

Statement 1 is a special case of the more general Statement 2 formulated below.
\medskip

\textit{Example 2}.
Let us consider another nonlinear equation of the reaction-diffusion type
\begin{equation}
u_t=a_1u_{xx}+u_x^2+bu^2+a_2\qquad (b>0),
\label{xeq12:05}
\end{equation}
where $a_1$, $a_2$, and $b$ are free parameters.
Equation \eqref{xeq12:05} admits an exact solution with generalized separation of variables containing a trigonometric function $x$ of the form
\begin{equation}
u=\psi_1(t)\cos(\sqrt b\,x)+\psi_2(t),
\label{xeq12:06}
\end{equation}
where the functions $\psi_1=\psi_1(t)$ and  $\psi_2=\psi_2(t)$
are described by the following nonlinear ODE system:
\begin{equation}
\begin{aligned}
&\psi_1'=2b\psi_1\psi_2-a_1b\psi_1,\\
&\psi_2'=b(\psi_1^2+\psi_2^2)+a_2.
\label{xeq12:07}
\end{aligned}
\end{equation}

Formally replacing the parameters $a_1$ and $a_2$ in equation \eqref{xeq12:05} with arbitrary functions $a_1(t)$ and $a_2(t)$, we obtain
\begin{equation}
u_t=a_1(t)u_{xx}+u_x^2+bu^2+a_2(t)\qquad (b>0).
\label{xeq12:08}
\end{equation}
It is easy to check that equation \eqref{xeq12:08} also admits an exact solution of the form~\eqref{xeq12:06},
where the functions $\psi_1=\psi_1(t)$ and $\psi_2=\psi_2(t)$ satisfy the nonlinear ODE system \eqref{xeq12:07},
in which the parameters $a_1$ and $a_2$ must be replaced by arbitrary functions $a_1(t)$ and $a_2(t)$.

An attempt to similarly replace the constant $b$ in equation \eqref{xeq12:05} with an arbitrary function $b(t)$ will be unsuccessful,
since in this case the function \eqref{xeq12:06} with $b=b(t)$ will now not be an exact solution to such a modified equation. This happened because
the coordinate function $\varphi(x)=\cos(\sqrt b\,x)$ in the solution \eqref{xeq12:06} explicitly depends on the parameter $b$.

The considered illustrative example allows us to make a fairly obvious generalization, which we formulate in the form of the following statement.
\medskip

\textbf{Statement 2}.
\textit{Let the autonomous nonlinear partial differential equation~\eqref{bla1*},
depending on the free parameters $a_1,\dots,a_p$, has a generalized separable solution of the form
	\begin{align}
		u=\sum^m_{k=1}\psi_k(t)\varphi_k(x),
		\label{bla4*}
	\end{align}
in which all linearly independent coordinate functions $\varphi_k(x)$ do not depend on the parameters $a_1,\dots,a_p$,
and the functions $\psi_k=\psi_k(t)$ are described by an autonomous ODE system,
	\begin{align}
		\psi_k'=f_k(\psi_1,\dots,\psi_m;a_1,\dots,a_p),\quad \ k=1,\dots,m.
		\label{bla5*}
	\end{align}
Then a more complex non-autonomous partial differential equation \eqref{bla3*}, which is obtained from equation \eqref{bla1*}
by formally replacing the parameters $a_1,\dots,a_k$ with arbitrary functions $a_1(t),\dots,a_p( t)$, admits an exact solution of the form~\eqref{bla4*},
where the coordinate functions $\varphi_k(x)$ do not change, and the functions $\psi_k=\psi_k(t)$ are described by a non-autonomous ODE system,
	\begin{align}
		\psi_k'=f_k\bigl(\psi_1,\dots,\psi_m;a_1(t),\dots,a_p(t)\bigr),\quad \ k=1,\dots,m.
		\label{bla6*}
	\end{align}}
\smallskip

This statement can be proven using the results of \cite{gal2007} (see also \cite{polzai2012,polzhu2022}).
Indeed, let us substitute \eqref{bla4*} into the autonomous equation \eqref{bla1*}, and then eliminate the time derivatives using ODEs \eqref{bla5*}.
Further replacing the functions $\psi_k(t)$ with arbitrary constants $C_k$, we obtain
\begin{equation}
F\,\Bl[\sum^m_{k=1}C_k\varphi_k(x)\Br]=
\sum^m_{k=1}f_k(C_1,\ldots,C_m;a_1,\dots,a_p)\varphi_k(x),
\label{eqg:118}
\end{equation}
where $F[u]$ denotes the right-hand side of equation \eqref{bla1*}.
Relations \eqref{eqg:118} mean (see \cite{gal2007,polzhu2022}) that a finite-dimensional linear subspace,
\begin{equation}
\mathcal{L}_{\,m}=\bl\{\varphi_1(x),\ldots,\varphi_m(x)\br\},
\label{eqg:119}
\end{equation}
the elements of which are all possible linear combinations of coordinate functions $\varphi_1(x)$, \dots, $\varphi_m (x)$ included in the solution \eqref{bla4*},
is invariant with respect to the nonlinear differential operator in the spatial variable~$x$, which is on the right side of the autonomous equation \eqref{bla1*}.
Nonlinear differential operator on the right side of the more complex non-autonomous partial differential equation \eqref{bla3*}
obtained from the equation \eqref{bla1*} by formally replacing the free parameters $a_1,\dots,a_k$ with arbitrary functions $a_1(t),\dots,a_p(t)$,
depends parametrically on time $t$ (contains derivatives only with respect to the variable~$x$).
Therefore, when substituting the expression \eqref{bla4*} into the right side of the equation~\eqref{bla3*}, the functions $a_1(t),\dots,a_p(t)$ behave like constants \cite{polzhu2022},
which ultimately and leads to the non-autonomous ODE system \eqref{bla6*} for functions $\psi_k=\psi_k(t)$.
\medskip

\textit{Remark 1}.
On the left side of the nonlinear PDEs \eqref{bla1*} and \eqref{bla3*}, instead of the first derivative $\,u_t$ there can be the second derivative $\,u_{tt}$
or any linear combination of derivatives with respect to $t$ of the form $\,L[u]=\sum^s_{i=1}c_i(t)u^{(i)}_t$.
In this case, in Statement 2 on the left side of the ODEs \eqref{bla5*} and \eqref{bla6*}, the first derivatives $\,\psi_k'$ must be replaced by the second derivatives
$\,\psi_k''$ or expressions $\,L [\psi_k]$.
\medskip

Statement 2 allows for a simplified, more concise (but less informative) formulation, which is given below.
\medskip

\textbf{Statement 2a}.
\textit{Let the autonomous partial differential equation \eqref{bla1*}, depending on the free parameters $a_1,\dots,a_p$,
has solution \eqref{bla4*}, in which the linearly independent coordinate functions $\varphi_k(x)$ are not depend on the parameters $a_1,\dots,a_p$.
Then the more complex non-autonomous partial differential equation \eqref{bla3*}, obtained from equation \eqref{bla1*}
by formally replacing the parameters $a_1,\dots,a_k$ with arbitrary functions $a_1(t),\dots,a_p(t )$ also admits an exact solution of the form \eqref{bla4*},
in which the coordinate functions $\varphi_k(x)$ do not change, and the functions $\psi_k=\psi_k(t)$ are described by an appropriate system of ODEs.}
\medskip

Statements 1 and 2 can also be generalized to the case of nonlinear PDEs, which allow more complex exact solutions
with functional separation of variables. Let us illustrate this with a specific example.
\medskip

\textit{Example 3}.
Let us consider a reaction-diffusion type equation with logarithmic nonlinearity,
\begin{equation}
u_t=a_1u_{xx}+a_2u\ln u,
\label{xeq12:09}
\end{equation}
where $a_1$ and $a_2$ are free parameters.
Equation \eqref{xeq12:09} admits a functional separable solution of the form \cite{gal2007,polzhu2022}:
\begin{equation}
u=\exp[\psi_1(t)x^2+\psi_2(t)x+\psi_3(t)],
\label{xeq12:10}
\end{equation}
where the functions $\psi_n=\psi_n(t)$ ($n=1,\,2,\,3$)
satisfy the nonlinear ODE system
\begin{equation}
\begin{aligned}
\psi_1'&=4a_1\psi_1^2+a_2\psi_1,\\
\psi_2'&=4a_1\psi_1\psi_2+a_2\psi_2,\\
\psi_3'&=a_2\psi_3+2a_1\psi_1+a_1\psi_2^2.
\label{xeq12:11}
\end{aligned}
\end{equation}

Now formally replacing the parameters $a_1$ and $a_2$ in equation \eqref{xeq12:09} with arbitrary functions $a_1(t)$ and $a_2(t)$, we obtain a more complex equation,
\begin{equation}
u_t=a_1(t)u_{xx}+a_2(t)u\ln u,
\label{xeq12:12}
\end{equation}
It is easy to check that equation \eqref{xeq12:12} also admits an exact solution of the form \eqref{xeq12:10},
where the functions $\psi_n=\psi_n(t)$ ($n=1,\,2,\,3 $) satisfy the nonlinear ODE system \eqref{xeq12:11},
in which the parameters $a_1$ and $a_2$ must be respectively replaced by arbitrary functions $a_1(t)$ and $a_2(t)$.
\medskip

The considered illustrative example allows us to make a fairly simple generalization given below.
\medskip

\textbf{Statement 3}.
\textit{Let the autonomous nonlinear partial differential equation~\eqref{bla1*},
depending on the free parameters $a_1,\dots,a_p$, has a functional separable solution of the form
	\begin{align}
		u=U(z),\quad \ z=\sum^m_{k=0}\psi_k(t)x^k,
		\label{bla7*}
	\end{align}
where $U(z)$ is some function independent of the parameters $a_1,\dots,a_p$.
Then the more complex non-autonomous partial differential equation \eqref{bla3*},
which is obtained from equation \eqref{bla1*} by formally replacing the parameters $a_1,\dots,a_p$ with arbitrary functions $a_1(t),\dots,a_p(t)$,
also admits an exact solution of the form \eqref{bla7*} with the same function $U(z)$, but with different functions $\psi_k(t)$.}
\medskip

This statement is proven by passing in the PDEs \eqref{bla1*} and \eqref{bla3*} to a new variable~$z$ using the substitution $u=U(z)$,
after which the transformed equations will have generalized separable solutions to which Statement 1 is already applicable.

Statement 3 allows for further generalization, described below.
\medskip

\textbf{Statement 4}.
\textit{Let the autonomous nonlinear partial differential equation~\eqref{bla1*},
depending on the free parameters $a_1,\dots,a_p$, has a functional separable solution of the form
	\begin{align}
		u(x,t)=U(z),\quad \hbox{where}\quad z=\sum^m_{k=1}\varphi_k(x)\psi_k(t),
		\label{bla8*}
	\end{align}
in which all coordinate functions $\varphi_k(x)$ and the external function $U(z)$ do not depend on the parameters $a_1,\dots,a_p$,
and the functions $\psi_k=\psi_k(t)$ are described by the autonomous ODE system \eqref{bla5*}.
Then the more complex non-autonomous partial differential equation \eqref{bla3*}, which is obtained from the equation \eqref{bla1*}
by formally replacing the parameters $a_1,\dots,a_p$ with arbitrary functions $a_1(t),\dots,a_p( t)$, admits an exact solution of the form \eqref{bla8*},
where the coordinate functions $\varphi_k(x)$ and the function $U(z)$ do not change, and the functions
$\psi_k=\psi_k(t)$ are described by the non-autonomous ODE system \eqref{bla6*}.}
\medskip

Statement 4 is proven by passing to a new variable $z$ in the PDEs \eqref{bla1*} and \eqref{bla3*} using the substitution $u=U(z)$, after which the transformed equations
will have generalized separable solutions to which Statement 2 applies.

Statement 4 allows for a simplified, shorter (but less informative) formulation, which is given below.
\medskip

\textbf{Statement 4a}.
\textit{Let the autonomous nonlinear partial differential equation~\eqref{bla1*},
depending on the free parameters $a_1,\dots,a_p$, has a functional separable solution of the form \eqref{bla8*},
in which all coordinate functions $\varphi_k(x)$ and the external function $U(z)$ do not depend on the parameters $a_1,\dots,a_p$.
Then the more complex non-autonomous partial differential equation \eqref{bla3*}, which is obtained from the equation \eqref{bla1*}
by formally replacing the parameters $a_1,\dots,a_p$ with arbitrary functions $a_1(t),\dots,a_p( t)$, admits an exact solution of the form \eqref{bla8*},
where the coordinate functions $\varphi_k(x)$ and the function $U(z)$ do not change, and the functions
$\psi_k=\psi_k(t)$ are described by an appropriate ODE system.}

\section{Using solutions of PDEs with constant delay to construct exact solutions of PDEs with variable delay}\label{sec-4}

Let us describe a method for constructing exact solutions of PDEs with variable delay of general form,
based on the use of solutions of related simpler PDEs with constant delay.
We will demonstrate the sequence and logic of useful preliminary reasoning in such cases using two specific examples,
after analyzing which we will formulate general conclusions in the form of statements.
\medskip

\textit{Example 4}.
Let us consider the reaction-diffusion equation with quadratic nonlinearity and constant delay
\begin{align}
	u_t=a(uu_x)_x+b\bar u,\quad \ \bar u=u(x,t-\tau),
	\label{xeq:01aa}
\end{align}
with $\tau=\text{const}>0$,
which admits a generalized separable solution of the form
\begin{align}
	u=\psi_1(t)x^2+\psi_2(t),
	\label{xeq:04aa}
\end{align}
where the functions $\psi_1=\psi_1(t)$ and  $\psi_2=\psi_2(t)$
are described by a nonlinear ODE system with constant delay,
\begin{equation}
\begin{aligned}
	&\psi_1'=6a\psi_1^2+b\bar\psi_1,\quad \ \bar\psi_1=\psi_1(t-\tau),\\
	&\psi_2'=2a\psi_1\psi_2+b\bar\psi_2,\quad \ \bar\psi_2=\psi_2(t-\tau).
\end{aligned}
\label{xeq:05aa}
\end{equation}

Replacing the constant $\tau>0$ in \eqref{xeq:01aa} with an arbitrary positive time function $\tau(t)$, we obtain a more complex nonlinear
reaction-diffusion equation with quadratic nonlinearity and variable delay,
\begin{align}
	u_t=a(uu_x)_x+b\bar u,\quad \ \bar u=u(x,t-\tau(t)).
	\label{xeq:06aa}
\end{align}

Since the nonlinear terms containing derivatives with respect to $x$ on the right-hand sides of equations \eqref{xeq:01aa} and
\eqref{xeq:06aa} are the same, it is natural to assume that the power structure of solutions in the spatial variable of both equations will also be the same,
and the time-dependent functional coefficients at various powers of~$x$ will change.

In other words, we are looking for exact solutions to the reaction-diffusion equation with variable delay \eqref{xeq:06aa}
in the same form \eqref{xeq:04aa} as solutions to the original equation with constant delay \eqref{xeq:01aa}.
As a result, for the functions $\psi_1=\psi_1(t)$, $\psi_2=\psi_2(t)$ we obtain the ODE system~\eqref{xeq:05aa},
in which the constant $\tau$ must be replaced by the function $\tau(t)$.
\medskip

The results described in this example allow for the generalization formulated below.
\medskip

\textbf{Statement 5}.
\textit{Let the nonlinear PDE with constant delay
	\begin{equation}
		u_t=F\bigl(x,u,u_x,\dots,u^{(n)}_x;\bar u,\bar u_x,\dots,\bar u^{(p)}_x\bigr),\quad \ \bar u=u(x,t-\tau),
		\label{bla1***}
	\end{equation}
where $n>p$, have the generalized separable solution
	\begin{align}
		u=\sum^m_{k=1}\varphi_k(x)\psi_k(t),
		\label{bla2***}
	\end{align}
in which the linearly independent coordinate functions $\varphi_k(x)$ do not depend on~$\tau$,
and the functions $\psi_k=\psi_k(t)$ are described by the ODE system with constant delay
	\begin{align}
		\psi_k'=f_k(\psi_1,\dots,\psi_m;\bar\psi_1,\dots,\bar\psi_m),\quad \ \bar\psi_k=\psi_k(t-\tau),\quad \ k=1,\dots,m.
		\label{bla3***}
	\end{align}
Then a more complex PDE with variable delay,
	\begin{equation}
		u_t=F\bigl(x,u,u_x,\dots,u^{(n)}_x;\bar u,\bar u_x,\dots,\bar u^{(p)}_x\bigr),\quad \ \bar u=u\bigl(x,t-\tau(t)\bigr),
		\label{bla4***}
	\end{equation}
which is obtained from equation \eqref{bla1***} by formally replacing the constant $\tau$ with an arbitrary time function $\tau(t)$,
also admits an exact solution of the form \eqref{bla2***},
where the coordinate functions $\varphi_k(x)$ do not change, and the functions $\psi_k=\psi_k(t)$ are described by an ODE system with variable delay,
	\begin{align}
		\psi_k'=f_k(\psi_1,\dots,\psi_m;\bar\psi_1,\dots,\bar\psi_m),\quad \ \bar\psi_k=\psi_k\bigl(t-\tau(t)\bigr),\quad \ k=1,\dots,m.
		\label{bla5***}
	\end{align}}

The proof of Statement 5 is carried out in the same way as the proof of Statement 2.
\medskip

\textit{Remark 2}.
On the left side of the nonlinear PDEs with delay \eqref{bla1***} and \eqref{bla4***}, instead of the first derivative $\,u_t$ there can be the second derivative $\,u_{tt}$
or any linear combination of derivatives with respect to $t$ of the form $\ ,L[u]=\sum^s_{i=1}a_i(t)u^{(i)}_t$.
In this case, in Statement 5 on the left side of the ODEs with delay \eqref{bla3***} and \eqref{bla5***}, the first derivatives $\,\psi_k'$
must be replaced by the second derivatives $\,\psi_k''$ or expressions $\,L [\psi_k]$.
\medskip

Statement 5 can be generalized to the case of nonlinear PDEs with delay, which allow more complex functional separable solutions.
Let us illustrate this with a specific example.
\medskip

\textit{Example 5}.
Let us consider the reaction-diffusion type equation with two logarithmic nonlinearities and constant delay
\begin{equation}
u_t=u_{xx}+u(a\ln^2u+b\ln \bar u),\quad \ \bar u=u(x,t-\tau),
\label{xeq12:09***}
\end{equation}
which for $a>0$ admits a functional separable solution of the form \cite{polzhu2013}:
\begin{equation}
u=\exp\bl[\psi_1(t)\varphi(x)+\psi_2(t)\br],\quad \ \varphi(x)=C_1\cos\bl(\sqrt{a}\,x\br)+C_2\sin\bl(\sqrt{a}\,x\br),
\label{xeq12:10***}
\end{equation}
where $C_1$ and $C_2$ are arbitrary constants, and the functions $\psi_1=\psi_1(t)$ and $\psi_2=\psi_2(t)$ satisfy the nonlinear ODE system with constant delay
\begin{equation}
\begin{aligned}
\psi_1'&=2a\psi_1\psi_2-a\psi_1+b\bar\psi_1,\quad \ \bar\psi_1=\psi_1(t-\tau),\\
\psi_2'&=a(C_1^2+C_2^2)\psi_1^2+a\psi_2^2+b\bar\psi_2,\quad \ \bar\psi_2=\psi_2(t-\tau).
\label{xeq12:11***}
\end{aligned}
\end{equation}

Now formally replacing the constant $\tau$ in the equation \eqref{xeq12:09***} with an arbitrary function $\tau(t)$, we obtain a more complex equation,
\begin{equation}
u_t=u_{xx}+u(a\ln^2u+b\ln \bar u),\quad \ \bar u=u\bl(x,t-\tau(t)\br).
\label{xeq12:12***}
\end{equation}
It is easy to check that equation \eqref{xeq12:12***} also admits an exact solution of the form \eqref{xeq12:10***},
where the functions $\psi_1=\psi_1(t)$ and $\psi_2=\psi_2(t)$
satisfy the nonlinear ODE system \eqref{xeq12:11***}, in which the constant $\tau$ must be replaced by an arbitrary function $\tau(t)$.
\medskip

The considered illustrative example allows us to make a generalization, which we will formulate in the form of a statement.
\medskip

\noindent\textbf{Statement 6}.
\textit{Let the nonlinear PDE with constant delay \eqref{bla1***} have a functional separable solution of the form
	\begin{align}
		u(x,t)=U(z),\quad \hbox{where}\quad z=\sum^m_{k=1}\varphi_k(x)\psi_k(t),
\label{new1}
	\end{align} in which the linearly independent coordinate functions $\varphi_k(x)$ and
the external function $U(z)$ are independent on $\tau$, and the functions
$\psi_k=\psi_k(t)$ are described by the ODE system with constant delay
\eqref{bla3***}. Then the more complex PDE with variable delay \eqref{bla4***}, which
is obtained from equation \eqref{bla1***} by formally replacing the constant $\tau$ with
an arbitrary function $\tau(t)$, admits an exact solution of the form \eqref{new1}, where
the coordinate functions $\varphi_k(x)$ and the function $U(z)$ are unchanged, and the
functions $\psi_k=\psi_k(t)$ are described by the ODE system with variable delay
\eqref{bla5***}.}
\medskip

Table~1 lists some nonlinear reaction-diffusion equations with variable delay, allowing
generalized and functional separable solutions.
It is believed that $\tau= \tau(t)$ is an arbitrary positive continuously differentiable function (in particular,
in the case of proportional delay in the equations one should put $\tau=(1-p)t$, i.e.\ $t-\tau=pt$ , where $0<p<1$).
Note that the functional separable solutions of the last two equations in Table~1 are presented in implicit form.

\begin{table}[!ht]
\textbf{Table 1.} Reaction-diffusion type equations with variable delay of a general
type, allowing solutions with generalized and functional separation of variables. Notation:
$\bar u=u(x,t-\tau(t))$, $\bar\psi=\psi(t-\tau(t))$; $f(z)$, $g(z)$, and $\tau(t)$ are arbitrary functions; $C_1$,
$C_2$, and $C_3$ are arbitrary constants. 

\medskip
\footnotesize
\def\vruleskip{\noalign{\nointerlineskip}
  \omit& height 2pt &\omit&&\omit&&\omit&\cr \noalign{\nointerlineskip}}
\def\rlineskip{\vruleskip\hline\vruleskip}
\tabskip=0pt
\halign to \textwidth{\strut#& \vrule#\tabskip=.2em plus1.8em&
  \hfil\vtop{\hsize=.26\textwidth \parindent=0pt \raggedright #\strut}\hfil& \vrule#&
  \hfil\vtop{\hsize=.41\textwidth \parindent=0pt \raggedright #\strut}\hfil& \vrule#&
  \hfil\vtop{\hsize=.27\textwidth \parindent=0pt \raggedright #\strut}\hfil& \vrule#\tabskip=0pt \relax\cr
\hline
\vruleskip
&& \strut Original equation&& \strut Form of exact solution&& \strut Determining equations&\cr
\vruleskip
\hline
\omit& height.5mm&\omit&&\omit&&\omit&\cr
\hline
\vruleskip && $u_t=au_{xx}+uf(\bar u/u)$&&
   $u=[C_1\cos(\beta x)+C_2\sin(\beta x)]\psi(t)$;\\
   $u=[C_1\exp(-\beta x)+C_2\exp(\beta x)]\psi(t)$;\\
   $u=(C_1x+C_2)\psi(t)$&&
   $\psi'_t=-a\beta^2\psi+\psi f(\bar\psi/\psi)$;\\
   $\psi'_t=a\beta^2\psi+\psi f(\bar\psi/\psi)$;\\
   $\psi'_t=\psi f(\bar\psi/\psi)$&\cr
\rlineskip && $u_t=au_{xx}+bu\ln u$\\ {}\qquad $+\,u f(\bar u/u)$&&
   $u=\varphi(x)\psi(t)$&&
      $a\varphi''_{xx}=C_1\varphi-b\varphi\ln\varphi$,\\
      $\psi'_t=C_1\psi+b\psi\ln\psi$\\ {}\qquad $+\,\psi f(\bar\psi/\psi)$&\cr
\rlineskip && $u_t=au_{xx}+f(u-\bar u)$&&
   $u=C_2x^2+C_1x+\psi(t)$&&
   $\psi'_t=2C_2a+f(\psi-\bar\psi)$&\cr
\rlineskip && $u_t=au_{xx}+bu$\\ {}\qquad $+\,f(u-\bar u)$&&
   $u=C_1\cos(\lambda x)+C_2\sin(\lambda x)+\psi(t)$,\\ {}\qquad where $\lambda=\sqrt{b/a}$ \ (if \ $ab>0$);\\
   $u=C_1\exp(-\lambda x)+C_2\exp(\lambda x)+\psi(t)$,\\ {}\qquad where $\lambda=\sqrt{-b/a}$ \ (if \ $ab<0$)&&
   $\psi'_t=b\psi+f(\psi-\bar\psi)$;\\
   \qquad\qquad\qquad\qquad\\
   $\psi'_t=b\psi+f(\psi-\bar\psi)$&\cr
\rlineskip && $u_t=a(u^ku_x)_x+uf(\bar u/u)$&&
   $u=\varphi(x)\psi(t)$&&
   $a(\varphi^k\varphi'_x)'_x=C_1\varphi$,\\
   $\psi'_t=C_1\psi^{k+1}+\psi f(\bar\psi/\psi)$&\cr
\rlineskip && $u_t=a(u^ku_x)_x+bu^{k+1}$\\ \qquad $+\,uf(\bar u/u)$&&
   $u=[C_1\cos(\beta x)+C_2\sin(\beta x)]^{\frac 1{k+1}}\psi(t)$,\\
   {}\quad \ \ where $\beta=\sqrt{b(k+1)/a\!}$, \ $b(k+1)>0$;\\
   $u=(C_1e^{-\beta x}+C_2e^{\beta x})^{\frac 1{k+1}}\psi(t)$,\\
   {}\qquad  where $\beta=\sqrt{-b(k+1)/a\!}$, $\,b(k+1)<\!0$;\\
   $u=C_1\exp\bigl(-\frac{b}{2a}x^2+C_2x\bigr)\psi(t)$\\
   {}\qquad  for $k=-1$;\\
   $u=\varphi(x)\psi(t)$ \ (generalizes previous\\ \qquad solutions)&&
   $\psi'_t=\psi f(\bar\psi/\psi)$;\\
   \qquad \\
   $\psi'_t=\psi f(\bar\psi/\psi)$;\\
   \qquad \\
   $\psi'_t=\psi f(\bar\psi/\psi)$;\\
   \qquad \\
   $a(\varphi^k\varphi'_x)'_x+b\varphi^{k+1}=C_1\varphi$,\\
   $\psi'_t=C_1\psi^{k+1}+\psi f(\bar\psi/\psi)$&\cr
\rlineskip && $u_t=a(e^{\lambda u}u_x)_x+f(u-\bar u)$&&
   $\ds u=\frac 1\lambda\ln(C_1\lambda x^2+C_2x+C_3)+\psi(t)$&&
      $\psi'_t=2aC_1e^{\lambda\psi}+f(\psi-\bar\psi)$&\cr
\rlineskip && $u_t=a(e^{\lambda u}u_x)_x+be^{\lambda u}$\\ \qquad $+f(u-\bar
u)$&&
   $u=\frac 1\lambda\ln[C_1\!\cos(\beta x)\!+\!C_2\sin(\beta x)]\!+\!\psi(t)$,\\
   {}\quad \ \ where $\beta=\sqrt{b\lambda/a}$, \ $b\lambda>0$;\\
   $u=\frac 1\lambda\ln (C_1e^{-\beta x}+C_2e^{\beta x})+\psi(t)$,\\
   {}\quad \ \ where $\beta=\sqrt{-b\lambda/a}$, \ $b\lambda<0$;\\
   $u=\varphi(x)+\psi(t)$ \ (generalizes previous\\ \qquad solutions)&&
   $\psi'_t=f(\psi-\bar\psi)$;\\
   \qquad \\
   $\psi'_t=f(\psi-\bar\psi)$;\\
   \qquad \\
   $a(e^{\lambda\varphi}\varphi'_x)'_x+be^{\lambda\varphi}=C_1$,\\
   $\psi'_t=C_1e^{\lambda\psi}+f(\psi-\bar\psi)$&\cr
\rlineskip
&& $u_t=[(a\ln u+b)u_x]_x$\\
   \qquad $-\,cu\ln u+uf(\bar u/u)$&&
   $\ds u=\exp(\pm \lambda x)\psi(t)$, \ $\lambda=\sqrt{c/a}$&&
      $\psi'_t\!=\lambda^2(a\!+b)\psi+\psi f(\bar\psi/\psi)$&\cr
\rlineskip
&& $u_t=a[f'(u)u_x]_x+b$\\
$\quad \ +\frac 1{f'(u)}g(f(u)-f(\bar u))$&&
   $\ds f(u)=\psi(t)-\frac{b}{2a}x^2+C_1x+C_2$&&
      $\psi'_t=g(\psi-\bar\psi)$&\cr
\rlineskip
&& $u_t=a[uf'(u)u_x]_x$\\
$\quad \ +\frac 1{f'(u)}[bf(u)+cf(\bar u)]$&&
   $\ds f(u)=\varphi(t)x+\psi(t)$&&
      $\varphi'_t=b\varphi+c\bar\varphi$,\\
      $\psi'_t=b\psi+c\bar\psi+a\varphi^2$&\cr
\vruleskip
\hline}
\end{table}

\section{Brief conclusions}

It has been established that generalized and functional separable solutions of nonlinear autonomous PDEs,
the coefficients of which do not depend on time $t$, can be used to construct exact solutions of more general nonlinear non-autonomous PDEs,
the coefficients of which depend arbitrarily on time. Examples of exact solutions of nonlinear non-autonomous PDEs are given.

It is shown that generalized and functional separable solutions of nonlinear PDEs
with constant delay make it possible to find exact solutions of more complex nonlinear PDEs with variable delay of general form.
Some nonlinear reaction-diffusion equations with variable delay are described, allowing exact solutions with a generalized separation of variables.


\renewcommand{\refname}{References}

\end{document}